\documentclass[10pt]{article}

\usepackage{amsfonts,amssymb,amsmath}
\usepackage{epsfig,chicago,float}
\usepackage[ansinew]{inputenc}
\usepackage[]{graphicx}

\renewcommand{\refname}{\centerline{REFERENCES}}
\begin{document}
\begin{center}
\textbf{\large Finding the centre: corrections for asymmetry in high-throughput sequencing datasets}

\vskip 0.25cm

\textbf{Jia R. Wu}$^{1}$\textbf{, Jean M. Macklaim}$^{1}$\textbf{, Briana L. Genge}$^{1}$\textbf{, and
Gregory B. Gloor}$^{1,2}$ \\
{\small $^{1}$Dep't of Biochemistry, U. Western Ontario, London, Canada, N6A 5C1\\
$^{2}$Dep't of Applied Mathematics, U. Western Ontario, London, Canada, N6A 5C1 \\
\textit{gbgloor@gmail.com} \\
}
\end{center}

\vskip 0.5cm {\centerline{\bf Abstract}}

High throughput sequencing is a technology that allows for the generation of millions of reads of genomic data regarding a study of interest, and data from high throughput sequencing platforms are usually count compositions. Subsequent analysis of such data can yield information on transcription profiles, microbial diversity, or even relative cellular abundance in culture. Because of the high cost of acquisition, the data are usually sparse, and always contain far fewer observations than variables. However, an under-appreciated pathology of these data are their often unbalanced nature: i.e, there is often systematic variation between groups simply due to presence or absence of features, and this variation is important to the biological interpretation of the data. A simple example would be comparing transcriptomes of yeast cells with and without a gene knockout. This causes samples in the comparison groups to exhibit widely varying centres. This work extends a previously described log-ratio transformation method that allows for variable comparisons between samples in a Bayesian compositional context. We demonstrate the pathology in  modelled and real unbalanced experimental designs to show how this dramatically causes both false negative and false positive inference. We then introduce several approaches to demonstrate how the pathologies can be addressed. An extreme example is presented where only the use of a predefined basis is appropriate. The transformations are implemented as an extension to a general compositional data analysis tool known as ALDEx2 which is available on Bioconductor. 


\vskip 1cm

\newpage

\section{Introduction}
\vskip-0.25cm High throughput sequencing (HTS) technology is used to generate information regarding the relative abundance of features.  In these designs, DNA or RNA is isolated, a library is made from a  sample of the nucleic acid, and a random sample of the library is sequenced on an instrument. The output is a set of short sequence tags, called reads, which are mapped to example sequences for each feature to generate a table of read counts per feature for every sample. Traditionally, samples comprise a set of features whose identity depends on the experimental design. For example, features are genes in the case of RNA-seq or  metagenomic sequencing, or are operational taxonomic units (OTUs) when the objective is identifying microbial diversity. 

These data are often analyzed by count based methods, such as negative binomial or zero-inflated Gaussian models, that assume the features are independent and identically distributed for statistical tests \shortcite{Auer:2010aa,Anders:2013aa}. However,  the capacity of the instrument used for HTS imposes an arbitrary upper limit on the total number of reads observed. Thus, data collected from high throughput sequencing are count compositions, and so counts per feature are not independent when collected in this way. In addition, several other pre- and post-sequencing steps contribute to make the data compositional \shortcite{gloorAJS:2016}. Traditional tools do not address the compositional nature of HTS data \shortcite{fernandes:2014,gloorAJS:2016} and assume that the features are sufficiently independent when there are enough of them, or when they fulfill certain statistical properties \shortcite{Weiss:2016aa}, although much effort is placed on `normalizing' the data to have a consistent read depth \shortcite{Sun:2013aa,McMurdie:2014a}.

Formally, \shortciteN{Aitchison:1986} defined a composition as a vector \textit{\textbf{x}}  of positive values \textit{x\textsubscript{1}}\ldots\textit{x\textsubscript{D}} whose features  sum to an arbitrary constrained constant $\alpha$. Absolute values of features in a composition are uninformative, and the only information provided in compositional data are the relative magnitudes of the ratios between the pairs of components. For example, the only knowledge available is that the gene 1:gene 2 ratio is 5, but the absolute abundance of either is unavailable.  Aitchison demonstrated that compositional data  can be properly analyzed by log-ratios between the features, since these data carry only relative information\shortcite{Aitchison:1986}.

One way of satisfying the need to examine the ratios between parts is to use the centred-log-ratio (CLR) transformation proposed by Aitchison, defined as: 

\begin{equation}
\begin{split}
\textbf{x}_{clr} &= log  \big( \frac{x_i}{g(\textbf{x})}   \big)_{i=1\dots D} \\
\text{where}~
	\textbf{x}_{clr} &= \text{A composition transformed by CLR} \\
	x_i &= \text{A feature of the non-transformed composition (\textbf{x})} \\
	D &= \text{The number of features of \textbf{x}} \\ 
	g(\textbf{x}) &= \text{Geometric mean of D features of \textbf{x}}
\end{split}
\label{eq:CLR}
\end{equation}

Since all arbitrary sums are the same this led to the concept of a composition as an equivalence class where composition \textit{\bf{x}} can be scaled into an identical composition \textit{\textbf{y}} by multiplication of a constant $\alpha$ \shortcite{barcelo:2001}. Thus, in the ideal case, we can discuss any composition as being a proportion scaled by $\alpha$ without loss of precision. Indeed, the CLR, and indeed any ratio-based method is, at least in theory, scale-invariant because if the parts of \textit{\textbf{x}} are counts with $\alpha=$\textit{N} reads, then: 

\begin{equation}
	\textbf{x}_{clr}= log\big( \frac{Nx_i}{g(N\textbf{x})}   \big) =  log\big( \frac{x_i}{g(\textbf{x})}  \big).
\label{eq:equip}
\end{equation}
The  important caveat that limits this ideal situation when dealing with high throughput sequencing data is that the total read count, $\alpha$, for each observation must be roughly similar.

Aitchison \citeyear{Aitchison:1986} also defined the ALR, the additive log-ratio as: 

\begin{equation}
\textbf{x}_{alr} = log  \big( \frac{x_i}{x_D}   \big)_{i=1\dots D-1}
\label{eq:ALR}
\end{equation}

where, following from above, $\textit{\textbf{x}}_{alr}$ is the composition transformed by ALR, and the denominator is the $D^{th}$ feature of \textit{\textbf{x}}, which by convention is the feature chosen to be constant.  

In the ALR, the log-ratio is thus determined by selecting one presumed invariant feature as the denominator. The ALR is surprisingly similar to the relative qPCR approach in common use in molecular biology that measures relative abundance of molecules in a mixture \shortcite{Thellin:1999aa,Vandesompele:2002aa}. Here, the feature of unknown abundance is determined relative to the abundance of a feature of (presumed) known abundance, which can be a housekeeping gene or can be a DNA molecule of known amount added to the mixture. It is well known that the relative abundance measure will change when a different DNA species is used as the denominator, leading to the use of multiple (presumed invariant) features in some cases. Thus, the ALR and CLR can be viewed as the two limits of a continuum of incomplete knowledge about the proper internal standard, or basis, by which relative abundance should be judged. The ALR uses one presumed constant feature as the basis; while the CLR  presumes that the majority of features are not changed, leading to the use of the geometric mean of all features as the basis. We can however, choose to use combinations of other features as the basis.

For convenience, the analyses and discussion here are drawn from RNA-seq, or transcriptome, experiments where the data are exploring the relative abundance of features that are gene transcripts found in cells in an environment. However, the examples, results and conclusions apply without restriction to metagenomic sequencing, microbial diversity sampling (by 16S rRNA gene sequencing) or to in-vitro selection experiments \shortcite{fernandes:2014}.

It is common for HTS data to be sparse, that is, for a given sample to contain features withcounts of 0. Furthermore, the sparsity of the samples is affected by the total number of reads obtained for each sample. Each sample in a transcriptome contains between thousands and tens of thousands of features each of which may have a potential dynamic range of over 4 orders of magnitude. In many cases a transcriptome dataset will be composed of several groups, where the expression of a feature (gene) is so low that it is below the detection limit in one group, and very high in another group. The expression of genes in biological systems is linked, and some genes control the expression of other genes, either by increasing or decreasing their relative abundance. Furthermore, the cell has a built-in control system whereby gene expression itself appears to be a composition, that is, the expression levels of all genes in a cell are constrained by an absolute upper bound  \shortcite{Scott:2010}. Note however, that this does not mean that a population of cells will have total gene expression with an upper bound, since the cells themselves can change in both absolute and relative abundance in a mixture. 

\section{Statement of the Problem}
\vskip-0.25cm

The assumption being made when using the CLR transformation to identify features that differ between groups is that most features are either invariant or varying at random when comparing the two groups. It is worth noting that this assumption is also made by essentially all differential abundance tools. This assumption is broken if there is any sort of systematic variation between groups. For example, when comparing microbial diversity between sampling sites or conditions, organisms present in one sub-site or condition may be absent from another \shortcite{Macklaim:2015aa,Hummelen:2010,Gajer:2012}. In the case of multi-organism RNA-seq (meta-RNA-seq), organisms resident in one condition may have a different expression profile and abundance than those resident in a second condition \shortcite{macklaim:2013}. In the case of a single-organism RNA-seq, samples from one condition may contain more genes than samples from another condition \shortcite{Lang:2015aa,Peng:2014aa,Zhao:2013aa,Gierlinski:2015aa}. These differences are represented by either zeroes or low count features that occur systematically in only one group. 

The potential for a change in cell number and the potential for expression linkage of genes in biological systems, coupled with the inability to collect a large enough number of sequence reads, can lead to experiments with an apparent or a real asymmetry in relative abundance of many genes or features. Such an asymmetry will result in mis-centering of the data when conducting differential abundance analyses, largely, but not exclusively because of the effect on the geometric mean upon which the CLR depends. The asymmetry will also affect the scale-invariance of the data, since a value of 0 is not scaled when multiplied by a constant. Note, that it is also entirely possible for the dataset \emph{as a whole} to be centred, but for the particular comparison of interest to not be centred. This could arise because of a systematic experimental bias that is unknown to the investigator.

\begin{figure}[ht]
\includegraphics[width=6in]{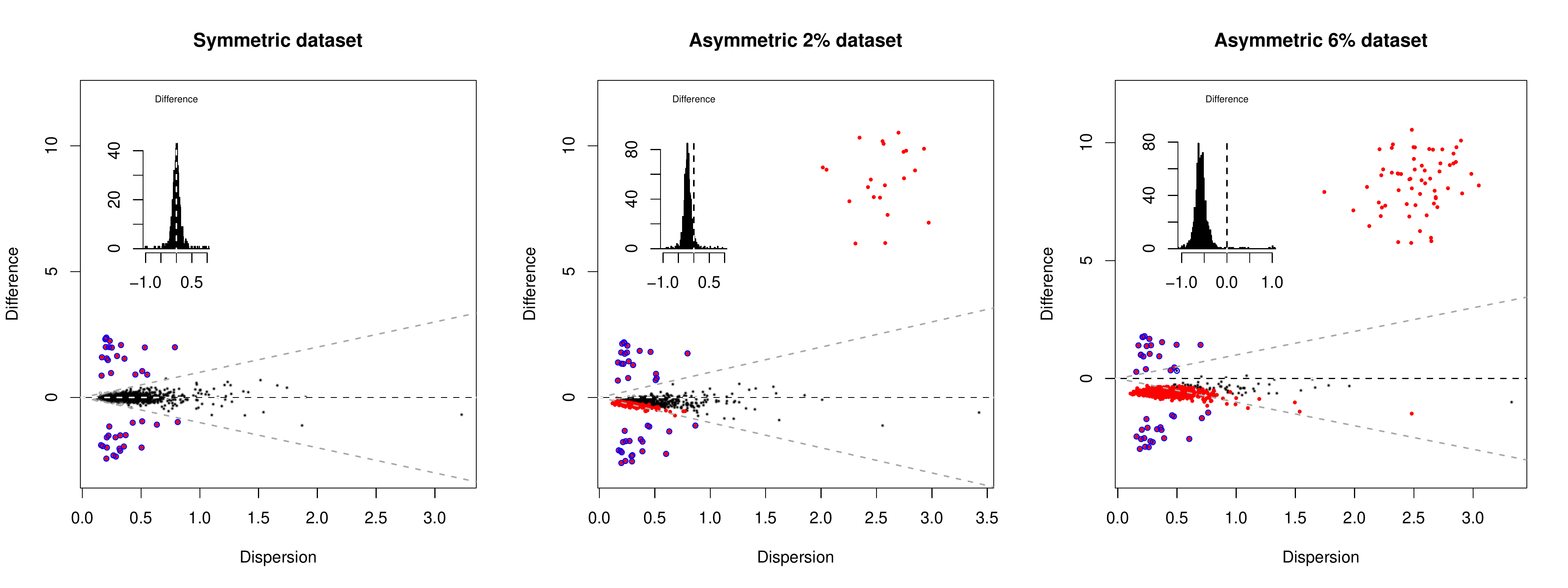}
\vspace{3mm} \caption{Effect plots  of simulated asymmetric data that illustrates the problem. The effect plots show the difference between two conditions in simulated RNA-seq data with 1000 genes where 40 genes are modelled to have true difference between groups. Each point is a feature (gene), they are coloured in black if not different between groups, red if identified as being statistically significantly different between groups, and red with a blue circle if they are one of the 40  genes modelled to be  true positives. The red points in the top right quadrant are the genes modelled to be asymmetrically variable between groups. These are also true positive features, but are not part of the initial modelled true positives. The inset histograms show the distribution of the differences between groups as calculated by ALDEx2, and the vertical line shows a difference of 0. These x-axis of these plots are truncated to show only differences near the midpoint.}
\label{Fig:f1a}
\end{figure}

Throughout, we use two plots to summarize the location of the features in multivariate datasets. The Bland-Altman (BA) plot  \shortcite{altman:1983} plots the mean log-ratio abundance on the x-axis and the difference between groups on the y-axis. The BA plot is efficient at showing the relationship between (relative, mean log-ratio) abundance and difference, but contains little information on the per-feature dispersion in the data. The Effect size plot \shortcite{Gloor:2015} complements the BA plot by showing the relationship between a measure of dispersion (on the x-axis) and the difference between groups (on the y-axis). All plots are in log units calculated a base of 2. The ratio between these two values is a proxy for the effect size statistic calculated by ALDEx2. Difference and dispersion are calculated using methods that are indifferent to distributional assumptions and are defined in the methods section. We also provide supplementary examples of the same data using compositional biplots \shortcite{aitchison2002biplots} to demonstrate that similar pathologies can occur when the data are displayed in this way. Figure \ref{Fig:f1a} shows that incorrect estimates of the location of the data can be achieved with seemingly minor variation within simulated data. The goal is to identify a basis  that best represents each sample so features can be accurately compared even when the data contains an asymmetry.

\section{Methods}\vskip-0.25cm

It is worth recalling that essentially all HTS data come from underpowered experimental designs, in the sense that there are more parts than there are samples: indeed it is common, because of cost to conduct and analyze only pilot-scale experiments. Thus, the strength of evidence for statistical inference must be weak, but paradoxically, the parts that are identified as differentially abundant must appear to be much more different between groups than the actual data support  \shortcite{Halsey:2015aa}. These can only be validated by replication or meta-analysis \shortcite{Cumming:2008aa}, both of which are rare in both the transcriptome and microbiome fields. 

When estimating differential abundance it is important to properly estimate the dispersion, $\tau$, of the $j^{th}$ feature for all samples; dispersion can be represented by the following simple model:

\begin{equation}
    \tau_{j} = \nu_j + \epsilon_j
\label{eq:dispersion}
\end{equation} 

where $\nu$ represents the underlying biological variation and $\epsilon$ represents the stochastic error  from all the steps involved in the collection, preparation, and sequencing of the dataset. The majority of extant analysis tools utilize point estimates of both parameters. First, it is generally assumed that  $\epsilon$ is small relative to $\nu$.  Second, it is assumed that there is some underlying similarity in the distribution of $\nu$ and $\epsilon$  for all features in all samples at a given relative abundance level. That is, if the $j$ features were ordered by abundance, that the expected value of $\nu_j$ would be  $\sim \sum (\nu_{j-m}\ldots \nu_{j+m}) / 2m $ where \textit{m} is some small offset in the abundance index. Similar logic applies to estimating the expected value of $\epsilon$, but many tools offer  more complex additional models to estimate these parameters for troublesome data. 

However, we  observed that $\epsilon$ can be exponentially larger than $\nu$ at the low count margin \shortcite{fernandes:2013,gloorAJS:2016}, and that properly accounting for this realization alone can result in an excellent fit to even problematic data. Thus, a reliable analysis can be obtained by incorporating an `in silico' technical replication which explicitly models the variation in $\epsilon$ as a probability density function on a per feature, per sample basis; in other words that $\tau_{j} = \nu_j + f(\epsilon_{j})$. This approach is implemented in the ALDEx2 Bioconductor package and substantially reduces the false positive identification rate in microbiome and transcriptome data while maintaining an acceptable true positive identification rate \shortcite{Thorsen:2016aa}.

The differences between groups, dispersion within groups and relative abundance were calculated using the ALDEx2 R package that uses Bayesian modelling that generates a probability function for  $\epsilon_j$ that can be used to place bounds on the uncertainty of the the observed data \shortcite{fernandes:2013,gloorAJS:2016}. If there are two groups, A and B,  this requires that the data comparison is properly centred on the difference between these groups. ALDEx2 has been shown to give meaningful and reproducible results, even on sparse, asymmetric datasets using many different experimental designs \shortcite{fernandes:2013,macklaim:2013,fernandes:2014,mcmurrough:2014}, although as shown here the asymmetry can still affect the outcome. 

I will adhere to the following notation when describing this process:
\begin{itemize}
\item{indices} will be denoted as lower case, italic; i.e., \textit{i,j,k,n}, except for the case of the number of features in a composition, when a \textit{D} will be used.
\item{a vector} will be denoted in bold, lower case, italic; i.e., the $i^{th}$ sample vector will be \textit{\textbf{s$_i$}}. Vectors derived from this vector will follow the same notation and contain $D$ features.
\item{a matrix or array} will be denoted in upper case, roman text; i.e, S
\end{itemize}

The starting point for analysis is an \textit{n} samples $ \times~D$ features  array. The  sample vector contains the number of reads mapped to any of the $j$  features in the $i^{th}$ sample,  $\textbf{s}_i=[j_1,j_2 \ldots j_D]$, where $i=1 \ldots n , j=1 \ldots D$. The total number of counts is irrelevant and determined by the machine \shortcite{Gloor:2016cjm,gloor2016s}. These data are compositional and are an example of an equivalence class with $\alpha_{i} = \sum \textbf{s}_{i}$. In theory, the vector $\textit{\textbf{s}}_i$ can be adjusted to a unit vector of proportions,  ${\textit{\textbf{p}}_i=[p_1,p_2 \ldots p_D] }$, i.e. $\alpha=1$, without loss of information by the maximum likelihood (ML) estimate  $\textit{\textbf{p}}_i=\textit{\textbf{s}}_i / \alpha_{i}$. In this representation, the value of the $j^{th}$ feature is a ML estimate of the probability of observing the counts conditioned on the fractional  $f$ that the feature represents in the underlying data and on the total read depth for the sample; i.e., $\mathbb{P}_{i,j}(f_{i,j}|\alpha_i)$. However, the maximum likelihood estimate will be exponentially inaccurate when the dataset contains many values near or at the low count margin \shortcite{Newey:1994} as is common in sparse HTS data. Instead we use a standard Bayesian approach \shortcite{Jaynes:2003} to infer a posterior distribution of the unit vector directly from $\textit{\textbf{s}}_i$, by drawing $k$ random Monte-Carlo instances from the Dirichlet distribution with a uniform, uninformative prior of 0.5, i.e.:

\begin{equation}
\textrm{P}_{i (1 \ldots k)}=
\left( \begin{array}{c}
    \textit{\textbf{p}}_1 \\
   \textit{\textbf{p}}_2 \\
    \vdots \\
    \textit{\textbf{p}}_k \\
\end{array} \right)=
\left( \begin{array}{ccccc}
    p_{i,11} & p_{i,21} & p_{i,31} & \dots  & p_{i,D1} \\
    p_{i,12} & p_{i,22} & p_{i,32} & \dots  & p_{i,D2} \\
    \vdots & \vdots & \vdots & \ddots & \vdots \\
    p_{i,1k} & p_{i,2k} & p_{i,3k} & \dots  & p_{i,Dk}\\
\end{array} \right)
\sim Dirichlet_{(1 \ldots k)}(\textit{\textbf{s}}_i + 0.5)
\label{eq:matrix}
\end{equation}

This approach has consistent sampling properties and removes the problem of taking a logarithm of 0 when calculating the CLR because the count 0 values are replaced by positive non-zero values that are consistent with the observed count data \shortcite{fernandes:2013,gloorAJS:2016}. Each of the Monte-Carlo instances, by definition, conserves proportionality and accounts for the fact that there is more information when $\alpha_i$ is large than when it is small. This partially restores scale invariance to the data by providing a distribution of values where the uncertainty of  features scales inversely with the read depth \shortcite{fernandes:2013,gloorAJS:2016}. 

Each of the Monte-Carlo Dirichlet instances are CLR-transformed row-wise using Equation \ref{eq:CLR} and the entire dataset is stored in the array $\textrm{C}$ with dimension ${n,D,k}$. Note that the CLR itself is scale invariant since the same output vector is obtained for all members of the equivalence class. For convenience a logarithm of 2 is used for the CLR transformation so that differences can be expressed in a intuitive scale.  

Summary statistics from the distribution of CLR values for each feature can be calculated and reported as either expected values or as medians of the distributions \shortcite{fernandes:2013}. If we have two groups, A and B, where the indices of the samples in the first group are $1 \ldots i_a$ and the indices of the samples in the second group are $i_{a + 1} \ldots n$, then the distributions of CLR values for the $j^{th}$ feature of the two groups can be contained in the vectors: $\textit{\textbf{a}}_j = \mathrm{C}_{(1 \ldots i_a)j(1...k)}$ and, $\textit{\textbf{b}}_j = \mathrm{C}_{(i_{a + 1} \ldots n)j(1...k)}$. Summary statistics use for plotting and analysis are: 

 
\begin{itemize}
\item{Log-ratio abundance} of a feature is the median of the joint distribution of CLR values from  groups A and B; i.e., it is the median of $ \textit{\textbf{a}}_j \cup \textit{\textbf{b}}_j$. 

\item{Dispersion} is the median of the vector $\Delta_{\textit{\textbf{a}}_j\lor \textit{\textbf{b}}_j}  = maximum(|\textit{\textbf{a}}_{j} - \textit{\textbf{a}}_{\langle j \rangle}|~,~|\textit{\textbf{b}}_{j} -\textit{\textbf{b}}_{\langle j \rangle}| )$, where $\langle j \rangle$ indicates a random permutation of the vector. The reported dispersion for each feature is denoted as $\tilde{\Delta}_{\textit{\textbf{a}}_j\lor \textit{\textbf{b}}_j}$ and is a conservative surrogate for the median absolute deviation when $\textit{\textbf{a}}_j$ and $\textit{\textbf{b}}_j$ contain many entries \shortcite{fernandes:2013}. 

\item{Difference} between groups is the median of the vector $\Delta_{\textit{\textbf{a}}_j-\textit{\textbf{b}}_j}  = (\textit{\textbf{a}}_{j} - \textit{\textbf{b}}_{\langle j \rangle})$, i.e., $\tilde{\Delta}_{\textit{\textbf{a}}_j-\textit{\textbf{b}}_j}$.

\item{Effect size} for a given feature is the median  the vector derived from $\Delta_{\textit{\textbf{a}}_j-\textit{\textbf{b}}_j} / \Delta_{\textit{\textbf{a}}_j\lor \textit{\textbf{b}}_j}$, and is thus a standardized difference between the distributions in $\textit{\textbf{a}}_{j}$ and $\textit{\textbf{b}}_{j}$.
\end{itemize}

\subsection{Simulated Data}
\vskip-0.25cm

RNA-Seq data was simulated for benchmarking purposes. Assemblies from \textit{Saccharomyces cerevisiae} uid 128 and  a complete reference genome of \textit{S. cerevisiae} were drawn from GenBank. The R package \texttt{polyester v1.10.0} \shortcite{polyester:2016} was used to simulate an RNA-Seq experiment with 2 groups of 10 replicates with 20x average sequencing coverage across the simulation experiment. For the base dataset, forty genes were chosen at random to have 2-5 fold expression difference, and these were apportioned equally between the two groups. These 40 features serve as an internal control of true positives for each dataset as their fold changes are explicit and should always be displayed as differentially expressed. We used bowtie2 \shortcite{bowtie2} to align the simulated reads  to the \textit{S. cerevisiae} reference genome. Labeling each group as A and B is arbitrary and hence the first 10 samples belong to condition A, and the final 10 sample belong to condition B. There are a total of 6349 features in these simulated data, but only the first 1000 genes by order were chosen for the majority of the figures. 

An additional 98 datasets derived from the base dataset are generated in order to benchmark how well the interquartile log-ratio (IQLR) transformation supports the assumption that most features are invariant and unchanging. As the original dataset has approximately 6000 nonzero features, 60 features are incrementally removed from the samples of condition A in each simulated dataset for a resultant set of datasets with sparsity ranging from 0\% to 98\% sparse in condition A. 

\begin{figure}[!ht]
\includegraphics[width=6in]{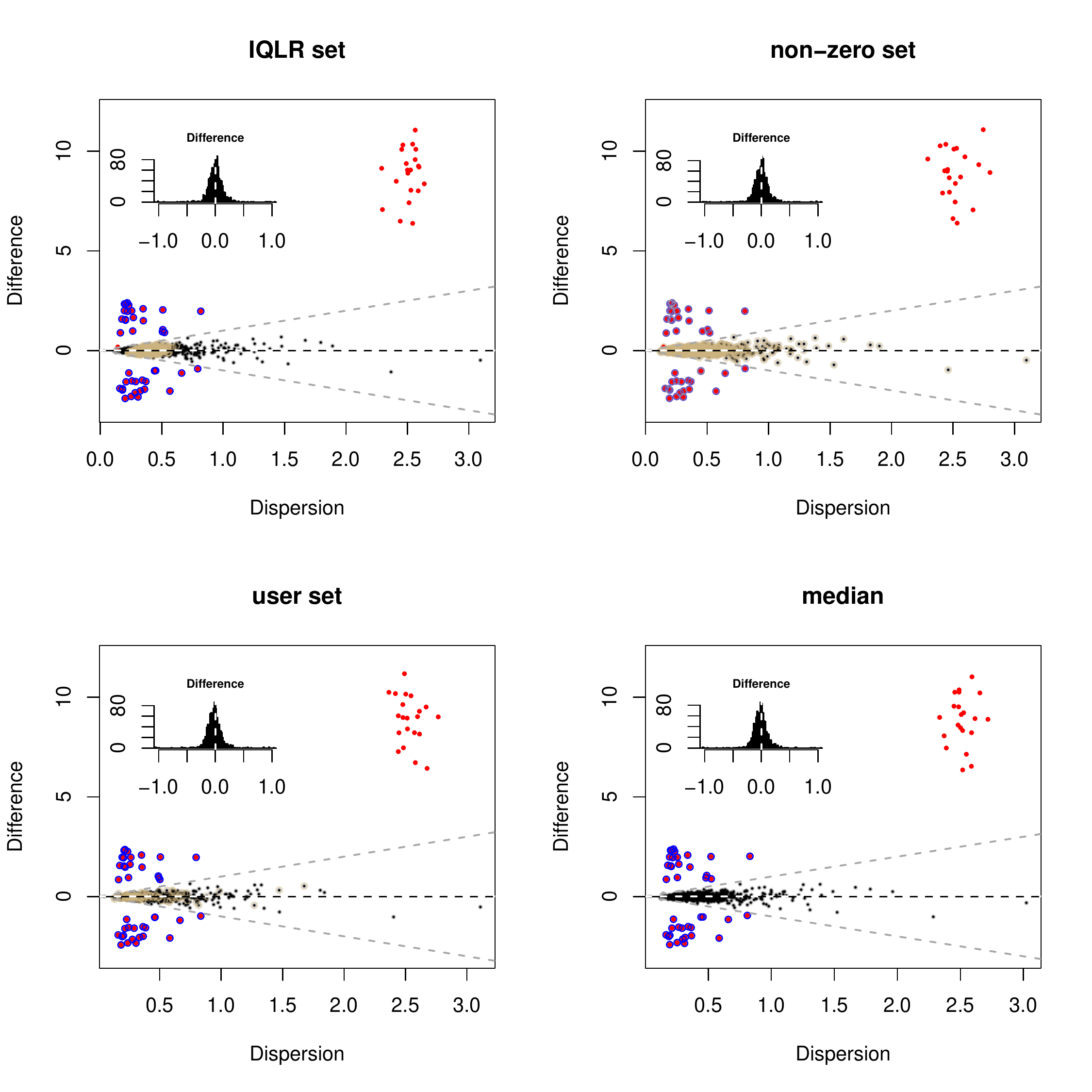}
\vspace{3mm} \caption{Effect plots of simulated asymmetric data with transformations that can result in a more accurate centring of the data. Points are coloured as in Figure \ref{Fig:f1a}, with the points used for the denominator in each case coloured in cyan.}
\label{Fig:f2a}
\end{figure}

\subsection{Four alternative methods}
\vskip-0.25cm

In its current implementation, ALDEx2 computes a per-sample geometric mean for the features and declares this as the baseline for feature comparisons. The `Symmetric dataset' panel in Figure \ref{Fig:f1a} is an effect plot demonstrating that the 40 internal control features are found to be both statistically significant and to have an effect size greater than 1 between the two groups, and the remainder of the features have very small difference, and correspondingly have an effect size much less than 1. The inset histogram shows the distribution of difference values between groups A and B, and it is clear that it is symmetric and has a location of 0. However, the introduction of small amounts of asymmetry strongly affect the results. The asymmetric 2\% dataset is the base dataset modified by setting the count value to 0 for 20 features chosen at random from Group A, and likewise the asymmetric 6\% dataset has 60 features from group A set to 0. It is apparent from the two right panels of Figure \ref{Fig:f1a} that this low level of simulated asymmetry breaks the assumption that most features are invariant, and the location of the difference between groups is no longer at the origin. Supplementary Figure 1 shows compositional PCA biplots of the same data, and here it is obvious that the centre of the data is not at the origin. Thus, the small amount of asymmetry is shifting the geometric mean of the data, causing bias. Thus, if even a proportion of features in a sample  do not follow the central tendency of the data, the geometric mean can be unreliable as a baseline. It is unlikely that the problem will be as easy to diagnose in real data as in simulated data. 

As can be seen in Equation \ref{eq:CLR}, the major determinant of the centre of a sample is the denominator, or basis, used to compute the CLR. Thus, one obvious approach to address the problem is to compute the geometric mean of a subset of features that are more representative of the central tendency of the data, and to use this value as the denominator in the equation. We  examined four different approaches to identifying the features to include in the denominator. 

The first approach was to identify those features that have variance which is most typical across all the samples. This was  done by calculating the variance of each feature after CLR transformation of the data, then identify  those features with a variance between the first and third quartiles of the dataset: this is referred to as the interquartile-variable feature \textit{IQVF} set of features. Thus, Equation 1 becomes:

\begin{equation}
IQLR_x = log  \big( \frac{x_i}{g(IQVF)}   \big)_{i=1 \dots D}
\label{eq:iqlr}
\end{equation}

where $IQLR_X$ is the transformed composition, and 	$g(IQVF)$ is the geometric mean of the IQVF features of X.

The transformation in Equation \ref{eq:iqlr} is termed the IQLR transformation. The results of this method are shown in Figure \ref{Fig:f2a}:IQLR on the Asymmetric 2\% dataset. The IQLR transformation restores the centre of the dataset to the origin, and the proper set of features is identified as being both significant, and having an effect size greater than 1.

The second approach uses as the denominator the set of non-zero features in each group.  Thus, in this case the geometric mean of group A and group B are based on different, but potentially overlapping, sets of features, this approach is called the no-zero log ratio (NZLR). As shown in Figure \ref{Fig:f2a}:non-zero, the NZLR method also restores the centre of the data to the origin and identifies the proper set of features as differential. 

\begin{figure}[th!]
\includegraphics[width=4in]{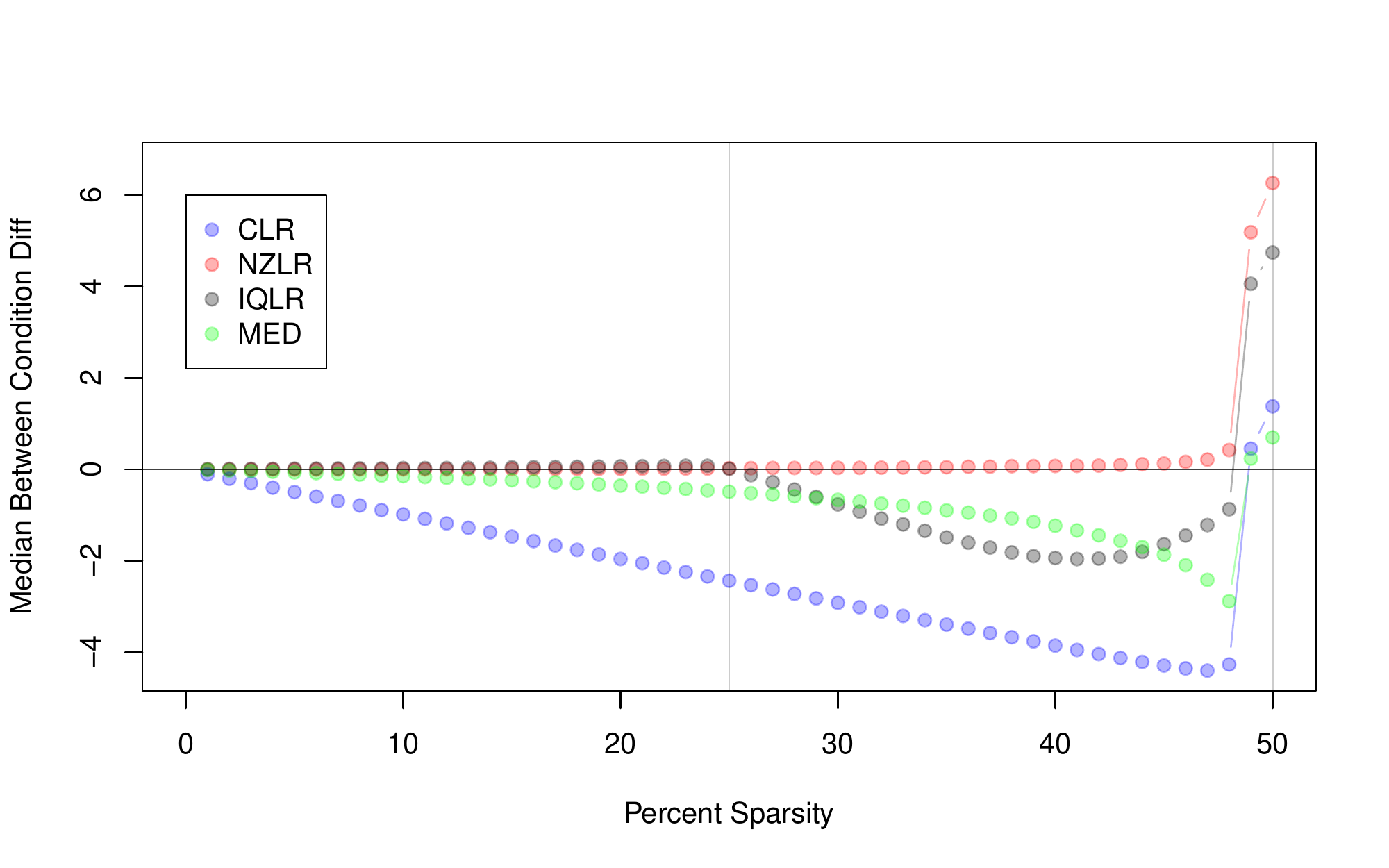}
\caption{The behaviour of each transformations for datasets with varying sparsity. Each point represents the median between condition difference for a given transformation in a dataset with a specified sparsity. Points closer to the location y=0 are favourable. The CLR transformation fails as soon as asymmetric sparsity is introduced. The IQLR  transformation is effective on datasets with up to 25\% asymmetric sparsity from zeroes or extreme count features. The NZLR transformation is effective on datasets with up to 50\% sparsity exclusively from zeroes. Replacing the geometric mean with the median in Equation \ref{eq:CLR}, is an improvement, but results in a generally small shift in midpoint.}
\label{Fig:failure}
\end{figure}

\begin{figure}[!t]
\includegraphics[width=5in]{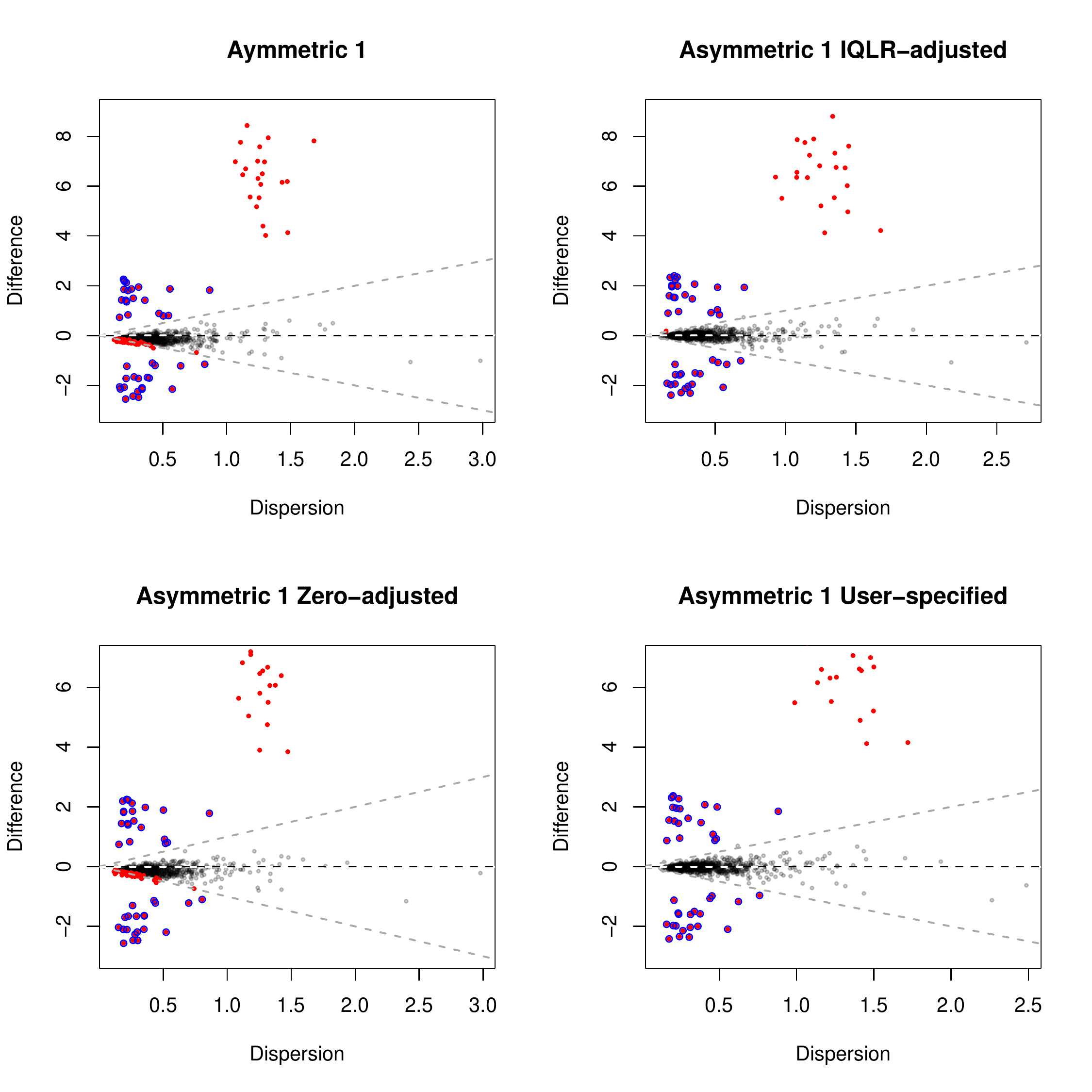}
\caption{The behaviour of each transformations for datasets with varying 2\% asymmetry where the asymmetric count value is 1. The top left panel shows that it is not sparsity that is the problem, but rather the asymmetry  between groups. An asymmetry caused by a count of 1 also induces false positive identifications, shown in red below the dashed line. Both the IQLR and the User-specified transformations are able to centre the data appropriately. However, the zero-adjustment method fails because the problem is not sparsity but low counts.}
\label{Fig:ones}
\end{figure}

The  third  approach uses as the denominator a set of user-defined features, and is termed the ULR for user-defined log-ratio. Thus, the user could choose to use one feature, in which case the approach would be the same as the ALR, or all features, in which case the result would be the same as the CLR, or a subset chosen based upon prior information. In the case of RNA-seq, this could include the set of genes involved in translation as these have been shown to be relatively stable across multiple conditions \shortcite{Scott:2010}, and can be presumed to represent a set of genes that are representative of the overall growth state of the cell. In principle, any set of features could be used although the investigator would need to present evidence for the  appropriateness of those features chosen in any particular experiment. In the example shown Figure \ref{Fig:f2a}, the `user set' also resulted in the location of the data being returned to 0.

The fourth approach replaces the geometric mean in Equation \ref{eq:CLR} with the  median since this should be a robust estimate of the midpoint of the data. 

\subsection{Limitations of the approaches}
\vskip-0.25cm
We  explored the limitations of these approaches in two ways. First, we examined how  sparsity affected the ability of the approaches to properly centre the data when dealing with asymmetric sparse data. Figure \ref{Fig:failure} shows that the centre of the CLR-transformed data deviates from 0 when the data have even very small amounts of asymmetric sparsity. The deviation is much smaller when the median is used as the denominator, but is not, in general, the best solution. Both the IQLR and NZLR approaches are able to properly centre the data when large amounts of asymmetric sparsity are present. The breakdown point for the IQLR method is 25\% sparsity in this dataset, and is approximately  45\% sparsity for the NZLR. Both, are obviously better choices that is the CLR, or the median choice when asymmetric sparsity is present.

Next, rather than modelling sparsity, we modelled low-count asymmetry by changing the asymmetry to a defined count of 1: note that any asymmetric count will behave similarly. In a biological context is entirely reasonable that asymmetry could occur because of low-counts  rather than sparsity. For example, the default gene expression condition for many genes is low-level expression, and the inclusion of a transcriptional activator could increase expression of many genes from very low expression to very high expression. In the context of 16S rRNA gene sequencing study, it is possible for samples to be dominated by one very abundant organism but to contain many other taxa at low abundance. Thus, we would have a low-count asymmetry that is not necessarily based on sparsity. Furthermore, sparsity is strongly affected by read depth, the same samples derived from a sequencing dataset from an Illumina NextSeq run delivering a total count of 400M reads will be substantially less sparse than those derived from an Illumina MiSeq run delivering a total count of 25M reads, however any underlying asymmetry will be preserved.

The results, shown in Figure \ref{Fig:ones} show that an asymmetry where the asymmetric value is 1 again results in the location of the data being displaced from 0. However, when the IQLR and ULR methods are used the location of the data is restored to 0. Not surprisingly, the NZLR does not restore the data to the proper location since the asymmetry is not driven by sparsity. The median method was not tested. We suggest that the NZLR method be used only when the other approaches fail, and when the investigator is confident that sparsity is driving the asymmetry in the data. 

\begin{figure}[ht]
\includegraphics[width=5in]{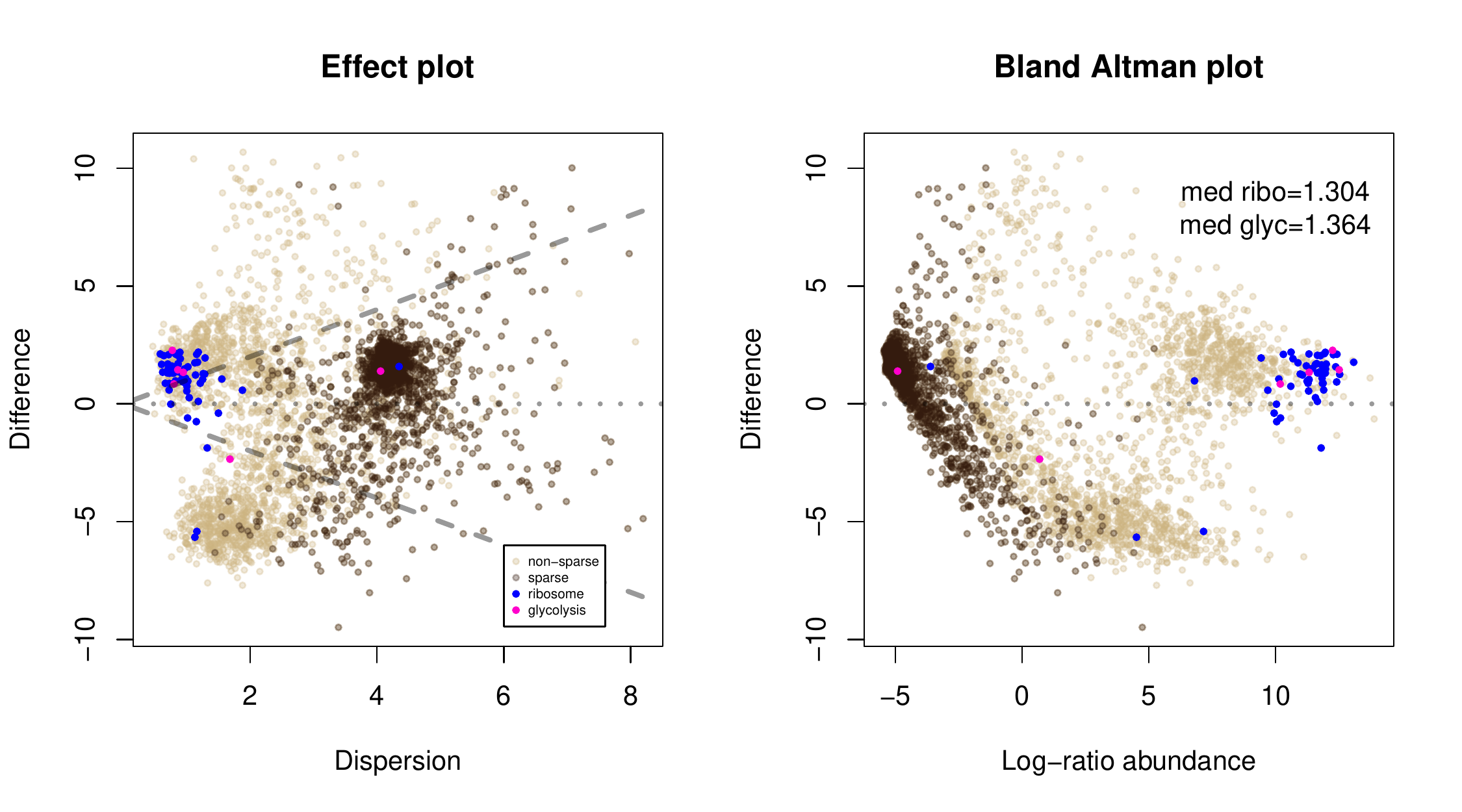}
\caption{Effect plot and Bland-Altman plots summarizing gene expression in two different states. }
\label{Fig:bv}
\end{figure}

\subsection{Example of a meta-RNA-seq dataset}
\vskip-0.25cm We finally introduce the example of an real RNA-seq dataset collected to determine the differences in gene expression of the vaginal microbial community in the healthy (H) and bacterial vaginosis (BV) states. The vaginal community can be dominated either by a few members of the \textit{Lactobacillus} genus in the H state, or by a mixed group of anaerobic bacterial genera in the BV state \shortcite{Ravel:2010}. In either state the members of the bacterial consortium from the other state are either very rare or absent. The data presented in Figure \ref{Fig:bv} show the distribution of between group difference, dispersion and relative abundance using an effect size plot  and a Bland-Altman (BA) plot. There are 10 H samples and 12 BV samples. Each point represents the intersect of the two given summary statistics taken from the ALDEx2 output for an individual protein or enzymatic function in the dataset \shortcite{macklaim:2013}. These values were computed using the CLR, and we can observe the pathologic nature of the data when using this log-ratio transformation on such an asymmetric and sparse dataset.

We can see that there is a large asymmetry in distribution, the most striking of which are the functions in BV located below the midline on the y-axis; there are a large number of features centred at about -2,-5 on the Effect plot and at 5,-5 on the BA plot. This asymmetry is driven by the greater complexity of the BV microbial community, and the generally larger and more complex genomes in the set of bacteria found in BV \shortcite{macklaim:2013}. The asymmetry is composed of both presence-absence (sparsity) and large differences between groups. This can be seen with the sparsity overlay color, where functions that contain one or more zeros are coloured in dark brown. However, note that there appears to be two clusters of functions that are just above the y-axis midline. These are composed of functions expressed at very low relative levels, that consequently fluctuate at the level of detection in the two groups, and are centred around 4,1.5 and -5,1.5 on the Effect and BA plots respectively. 

We also see a group composed of functions found in common between the the H and BV group that is  expressed at very high relative levels, centred around 1,1.5 and 8,1.5 on the two plots. These are functions that are central and required by all living organisms. Two sets of functions are highlighted: ribosomal protein functions in blue, and glycolytic functions in magenta. Many of the functions in these groups are often used as internal standards for comparison as it is assumed that their expression is invariant \shortcite{Scott:2010}. The median offset of the two sets on the y-axis is given, and we can see that they are both observed to be substantially above the expected location of 0 when the CLR is used to determine differential abundance.   

\begin{figure}[ht]
\includegraphics[width=5in]{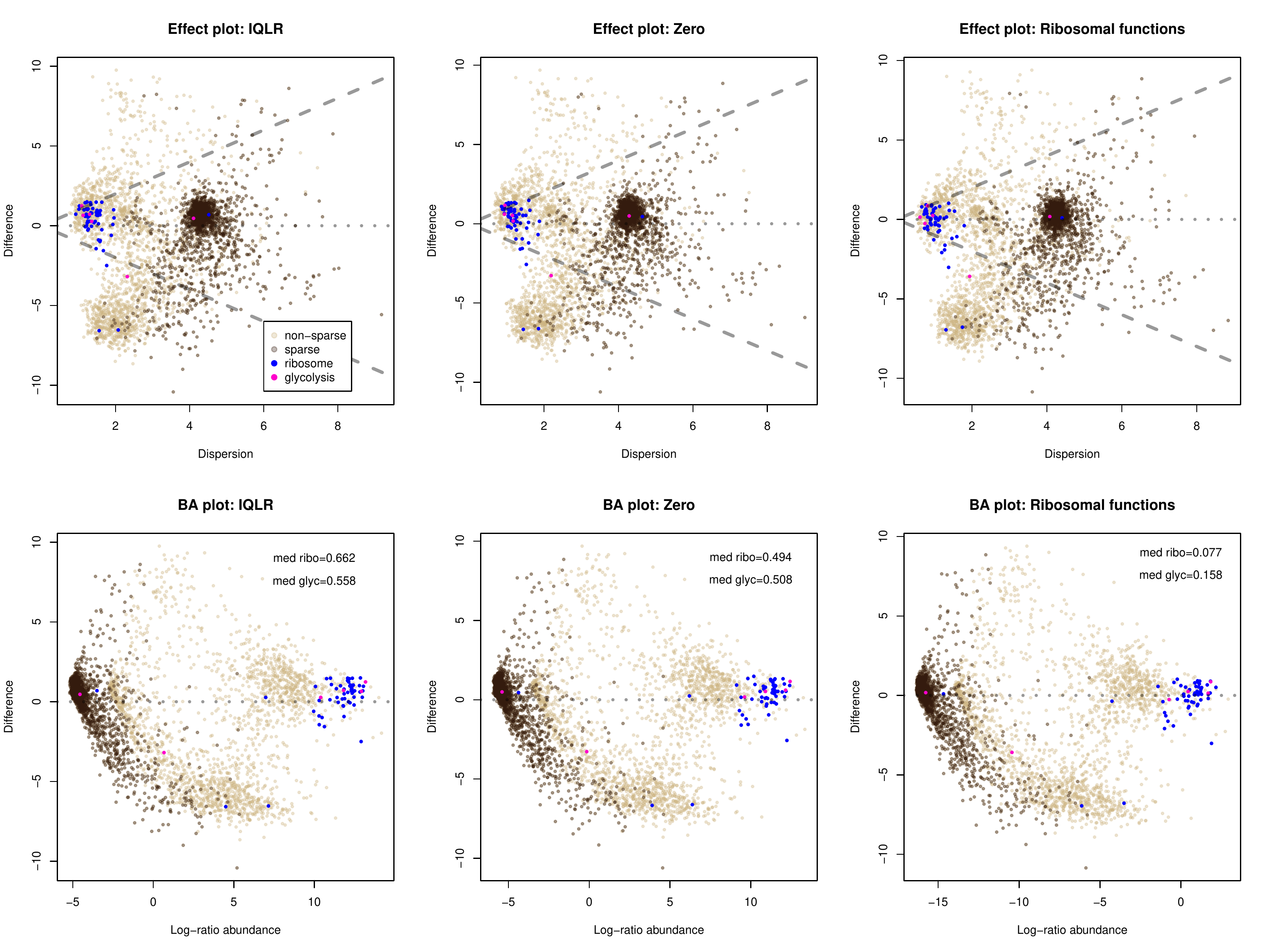}
\caption{Effect plot and Bland-Altman plots summarizing gene expression in two different states. }
\label{Fig:ribo}
\end{figure}

Finally, Figure \ref{Fig:ribo} shows the result of applying the three adjustment approaches to the asymmetric meta-RNA-sea dataset. Both the IQLR and zero-adjustment methods centred the data somewhat better than when the CLR was used. We can see that the bulk of the expected invariant groups are closer to the y-axis midline, and the offset of the ribosomal and glycolytic functions is reduced by half or more when compared to the CLR result. The zero-adjustment  appears to be slightly better than the IQLR method in this dataset, likely because of the large amount of asymmetric sparsity. However, it is clear that the asymmetry is so extreme that neither unsupervised approach is appropriate. Centring on the geometric mean of the ribosomal functions provides a substantial improvement. Here the very rare and assumed invariant functions appear to be near the y-axis centre line. It would be a tautology to test the appropriateness using the ribosomal functions, but we can see that the glycolytic functions are nearly centred, being only slightly above the midline.

\section{Conclusions}
\vskip-0.25cm

Biological data derived from high-throughput sequencing is rarely ideal and exhibits many pathologies. In particular, such data can be derived from asymmetric environments, where sets of genes, operational taxonomic units, or organisms can be present or abundant in one condition and absent or rare in another. Alternatively, an asymmetry in the data can arise because of a systematic failure in experimental design, for example, through improper blocking or the presence of outlier samples. In any of these instances the presence of an asymmetry may not be obvious. 

We demonstrated that even a small number of asymmetric features can change the location of the dataset, leading to both false positive and false negative differences being identified. We showed that the asymmetry can be associated with sparsity or by differences near the margin; in either case, the pathology was similar. We tested four different methods to properly centre the data, and found that the IQLR and user-specified centring approaches were the most general purpose and recommended for use, although all are implemented in the ALDEx2 R package. 

When the asymmetry is moderate, the IQLR correction is most appropriate. This correction makes the assumption that those features with variance that is found between the first and third quartile of variance, are a suitable proxy for the expected `typical' variance of the data. This approach can tolerate up to 25\% asymmetry in the data when the geometric mean of these features are used as the denominator in a log-ratio normalization. In fact, we recommend that the IQLR be used as the default when performing differential abundance analysis, since this normalization makes no strong assumption about the data and appears to never perform worse than the CLR normalization.

More extreme asymmetry, as found in our vaginal transcriptome dataset, forces the investigator to make strong assumptions about the underlying data. These assumptions are similar to the assumptions made when performing qPCR: that there are one or more invariant features in the data. We showed that making the assumption that features encoding common core metabolic functions in either information processing and translation, or glycolysis, behave similarly and can be used as a exemplars of `invariant' features. 

In either case, it must be remembered that the results of any analysis must be interpreted as \emph{abundance relative to the chosen invariant part of the dataset}, and not as changes in absolute abundance.

\section*{Acknowledgements and appendices}
\vskip-0.25cm

\renewcommand\refname{}
\vskip-1cm
\bibliography{Gloor_cw17_paper.bbl}

\end{document}